\documentclass[prd,preprint,superscriptaddress,showpacs,byrevtex]{revtex4} 
\usepackage{epsfig} 
\newcommand{\be}{\begin{equation}} 
\newcommand{\ee}{\end{equation}} 
\newcommand{\bea}{\begin{eqnarray}} 
\newcommand{\eea}{\end{eqnarray}} 

\begin{document} 

\begin{flushright}
YITP-SB-06-61
\end{flushright}

\title{ Schwinger Mechanism for Fermion Pair Production
in the Presence of Arbitrary Time 
Dependent Background Electric Field }

\author{Fred Cooper} \email{fcooper@nsf.gov} 
\affiliation{Physics Division, National Science Foundation, Arlington VA 22230}

\author{Gouranga C. Nayak} \email{nayak@insti.physics.sunysb.edu} 
\affiliation{ C. N. Yang Institute for Theoretical Physics, Stony Brook University, SUNY, Stony Brook, 
NY 11794-3840, USA } 

\date{\today} 

\begin{abstract} 
We study the Schwinger mechanism for the pair production of fermions in the
presence of an arbitrary time-dependent background electric field $E(t)$
by directly evaluating the path integral.
We obtain an exact non-perturbative result for the probability 
of fermion-antifermion pair production per unit 
time per unit volume per unit transverse momentum 
(of the fermion or antifermion) from the arbitrary time dependent electric 
field $E(t)$ via Schwinger mechanism. We find that the exact non-perturbative 
result is independent of all the time derivatives $\frac{d^nE(t)}{dt^n}$, 
where $n=1,2,....\infty$. This result has the same functional dependence on $E$
as the Schwinger's constant electric field $E$ result with the replacement: 
$E~\rightarrow~ E(t)$.

\end{abstract} 


\maketitle 

\newpage 
Using  a proper time method Schwinger derived the following 
one-loop non-perturbative formula 
\bea
\frac{dW}{d^4x}=\frac{e^2E^2}{4\pi^3} ~\sum_{n=1}^{\infty}  ~\frac{1}{n^2}~e^{-\frac{n\pi m^2}{|eE|}}
\eea
for the probability of $e^+e^-$ pair production per unit time per unit volume 
from a constant electric field E via vacuum polarization \cite{schw}. 
The $p_T$ distribution of the $e^+$ (or $e^-$) production, 
$\frac{dW}{d^4x d^2p_T}$, can not be obtained by using proper time method 
and a WKB approximate
method was used for this purpose in \cite{Casher}. A path integral method
can also be used to obtain the $p_T$ distribution \cite{gouranga}. In the case 
of fermions in QED one finds \cite{Casher}
\bea
\frac{dW}{d^4xd^2p_T}=-\frac{|eE|}{4\pi^3} {\rm Log}[1-e^{-\pi \frac{p_T^2+m^2}{|eE|}}].
\label{dsf}
\eea

Recently  this approach has been extended to QCD in SU(3)
gauge group by directly evaluating the path integral and the $p_T$ 
distribution of the quark and gluon production rate are found to 
depend on two independent Casimir invariants: $C_1=[E^aE^a]$ and
$C_2=[d_{abc}E^aE^bE^c]^2$ where $E^a$ is the constant chromo-electric 
field with color index a=1,2,..8 \cite{gouranga}. 

In a recent paper \cite{fredgouranga} we studied, for the first time, the
Schwinger mechanism in the presence of arbitrary time dependent electric 
field $E(t)$ by using a newly derived shift theorem \cite{shift} for the case of scalar field theory.
In this paper we will extend this study to fermions.

We obtain the following non-perturbative 
formula for the probability of fermion-antifermion 
pair production per unit time per unit volume per unit transverse 
momentum (of the fermion or antifermion) production from arbitrary 
time dependent electric field $E(t)$:
\bea
\frac{dW}{d^4xd^2p_T}=-\frac{|eE(t)|}{4\pi^3} {\rm Log}[1-e^{-\pi \frac{p_T^2+m^2}{|eE(t)|}}].
\label{dw}
\eea

This result has the remarkable feature that no time derivatives 
$\frac{d^nE(t)}{dt^n}$ (where n=1,2,...$\infty$) occur in the final 
expression. This result has the same functional dependence on $E$ as 
the Schwinger's constant  electric field $E$ result with the replacement:  
$E~\rightarrow~ E(t)$.

Now we present a derivation of eq. (\ref{dw}).

The Dirac Lagrangian density for fermion in the presence of a
background electro-magnetic field $A_{\mu}(t,x,y,z)$ is given by
\be
{\cal{L}}
=\bar{\psi} [(\hat{{p}\!\!\!\slash} -e{{A}\!\!\!\slash}) -m]
\psi =\bar{\psi} M[A] ~\psi
\label{laq}
\ee
where $\hat{p}_\mu=\frac{1}{i} \frac{\partial}{\partial x^\mu}$.
The vacuum to vacuum transition amplitude in the presence of background
field $A$ is given by
\bea
<0|0>^A=\frac{\int [d\bar{\psi}][d\psi] e^{i\int d^4x \bar{\psi} M[A] \psi}}{
\int [d\bar{\psi}][d\psi] e^{i\int d^4x \bar{\psi} M[0] \psi}} 
=Det[M[A]]/Det[M[0]] =e^{iS^{(1)}}.
\eea
The one loop effective action becomes
\bea
S^{(1)}= -i{\rm Tr} {\rm ln}
[(\hat{{p}\!\!\!\slash} -e{{A}\!\!\!\slash}) -m]
+i {\rm Tr} {\rm ln} [\hat{{p}\!\!\!\slash}  -m].
\label{efgl}
\eea
Since the full trace Tr is invariant under transposition we also get
\bea
S^{(1)}=-i{\rm Tr} {\rm ln}
[(\hat{{p}\!\!\!\slash} -e{{A}\!\!\!\slash}) +m]
+i {\rm Tr} {\rm ln} [\hat{{p}\!\!\!\slash}  +m].
\label{efg2}
\eea
Adding eqs. (\ref{efgl}) and (\ref{efg2}) we find
\bea
2S^{(1)}=-i{\rm Tr} {\rm ln}
[(\hat{{p}} -e{{A}})^2 +
\frac{e}{2}\sigma^{\mu \nu}F_{ \mu \nu}
-m^2] +i {\rm Tr} {\rm ln} [\hat{{p}}^2  -m^2]
\label{efg4}
\eea
where the Dirac tensor $\sigma^{\mu \nu}$ is given in terms of Dirac $\gamma$ matrices as follows:
\bea
\sigma^{\mu \nu} =\frac{i}{2}[\gamma^\mu \gamma^\nu - \gamma^\nu \gamma^\mu ].
\label{sig}
\eea
Since it is convenient to work with the trace of the exponential we use
\bea
{\rm ln} \frac{a}{b}=\int_0^\infty \frac{ds}{s} [
e^{-is(b-i\epsilon)} -e^{-is(a-i\epsilon)}].
\label{5}
\eea
Hence eq. (\ref{efg4}) becomes
\bea
2S^{(1)}=i {\rm Tr}  \int_0^\infty \frac{ds}{s} 
[e^{-is[(\hat{p} -eA)^2 + \frac{e}{2}\sigma^{\mu \nu}F_{ \mu \nu} -m^2-i\epsilon]} -e^{-is[\hat{p}^2  -m^2-i\epsilon]}].
\label{6}
\eea
The trace Tr is given by
\bea
{\rm Tr}={\rm tr}_x {\rm tr}_{\rm Dirac}
\label{tr}
\eea
where
\bea
{\rm tr}_x {\cal O}=\int d^4x <x| {\cal O} |x>.
\label{4c}
\eea

We assume that the time dependent electric field E(t)
is along the z-axis. We choose the Axial gauge $A_3=0$  
so that  only
\bea
A_0 =- E(t)z
\label{7}
\eea
is non-vanishing. Using eqs. (\ref{tr}), (\ref{4c}) and (\ref{7}) in eq. (\ref{6}) we obtain
\bea
 &&  S^{(1)}=\frac{i}{2} {\rm Tr_{Dirac}}  \int_0^\infty \frac{ds}{s} 
\int_{-\infty}^{+\infty} dt <t| \int_{-\infty}^{+\infty} dx 
<x| \int_{-\infty}^{+\infty} dy <y| \int_{-\infty}^{+\infty} dz <z| \nonumber \\
&&~ [ e^{-is[(\hat{p}_0+eE(t) z)^2-\hat{p}_z^2-\hat{p}_T^2-m^2+i\gamma^0 \gamma^3 E(t)-i\epsilon]} -
e^{-is(\hat{p}^2-m^2-i\epsilon)}] |z> |y> |x> |t>.
\label{10}
\eea
Inserting complete set of $|p_T>$ states $\int d^2 p_T |p_T><p_T|~=~1$ 
we find (we use the normalization $<q|p>=\frac{1}{\sqrt{2\pi}} ~e^{iqp}$)
\bea
&& S^{(1)}=\frac{i }{2(2\pi)^2} {\rm Tr_{\rm Dirac}}
\int_0^\infty \frac{ds}{s}  \int d^2x_T \int d^2p_T
e^{is(p_T^2+m^2+i\epsilon)}[\int_{-\infty}^{+\infty} dt <t|  
\int_{-\infty}^{+\infty} dz <z| \nonumber \\
&& e^{-is[(-i\frac{d}{dt}+eE(t) z)^2-\hat{p}_z^2+i\gamma^0 \gamma^3 E(t)]} |z> |t>- \int dt \int dz 
\frac{1}{4\pi s}].
\label{1p1}
\eea

Unlike the constant electric field calculation \cite{itzy, gouranga}, the Dirac trace in the above equation
can not be performed outside the "$t$" trace because the Dirac matrices are now multiplied by the time 
dependent electric field $E(t)$. The Dirac matrix \cite{itzy1}
\bea
\gamma^0 \gamma^3 = ~\alpha^3 =\left [ \begin{array}{cc}
0 & \sigma^3 \\
\sigma^3 & 0
\end{array} \right ]~=~
\left [ \begin{array}{cccc}
 0 & 0 & 1 & 0 \\
 0 & 0 & 0 & -1 \\
 1 & 0 & 0 & 0 \\
 0 & -1 & 0 & 0 
\end{array} \right ]
\eea
is symmetric and hence it can be diagonalized by an orthogonal matrix $U$:
\bea
(\gamma^0 \gamma^3) =U [(\gamma^0 \gamma^3)_{\rm Diagonal}] U^{-1}.
\eea
To perform the Dirac trace in eq. (\ref{1p1}) we 
diagonalize the $\gamma^0 \gamma^3$ matrix and find:
\bea
(\gamma^0 \gamma^3)_{\rm Diagonal} =
\left [ \begin{array}{cccc}
 \lambda_1 & 0 & 0 & 0 \\
 0 & \lambda_2 & 0 & 0 \\
 0 & 0 & \lambda_3 & 0 \\
 0 & 0 & 0 & \lambda_4
\end{array} \right ] ~=~
\left [ \begin{array}{cccc}
 1 & 0 & 0 & 0 \\
 0 & 1 & 0 & 0 \\
 0 & 0 & -1 & 0 \\
 0 & 0 & 0 & -1 
\end{array} \right ].
\label{dig}
\eea
Hence we find from eq. (\ref{1p1})
\bea
&& S^{(1)}=\frac{i }{2(2\pi)^2} \sum_{j=1}^4
\int_0^\infty \frac{ds}{s} \int d^2x_T \int d^2p_T
e^{is(p_T^2+m^2+i\epsilon)}[\int_{-\infty}^{+\infty} dt <t|  
\int_{-\infty}^{+\infty} dz <z| \nonumber \\
&& e^{-is[(-i\frac{d}{dt}+eE(t) z)^2-\hat{p}_z^2+i\lambda_j E(t)]} |z> |t>- \int dt \int dz 
\frac{1}{4\pi s}].
\label{11}
\eea

At this stage we use the shift theorem \cite{shift} 
\bea
&& \int dy \int_{-\infty}^{+\infty} dx f_1(y)<x|~e^{-[(a(y)x+\frac{d}{dy})^2+b(\frac{d}{dx})+
c(y)]}
~|x>f_2(y) \nonumber \\
&&= \int dy \int_{-\infty}^{+\infty} dx f_1(y)<x
-\frac{1}{a(y)}\frac{d}{dy}|~e^{-[a^2(y)x^2+b(\frac{d}{dx})+c(y)]} ~|x
-\frac{1}{a(y)}\frac{d}{dy}>f_2(y) \nonumber \\
\eea
to obtain:
\bea
&& S^{(1)}=\frac{i}{2(2\pi)^2}  \sum_{j=1}^4 \int_0^\infty\frac{ds}{s} \int d^2x_T\int d^2p_T
e^{is(p_T^2+m^2+i\epsilon)} \nonumber \\
&& [ \int_{-\infty}^{+\infty} dt <t| \int_{-\infty}^{+\infty} dz <z+\frac{i}{E(t)}\frac{d}{dt}| 
 e^{-is[e^2E^2(t) z^2-\hat{p}_z^2 +i\lambda_j E(t)]} |z+\frac{i}{E(t)}\frac{d}{dt}>|t> - \int dt  \int dz  \frac{1}{4\pi s}] \nonumber \\
\label{12a}
\eea
where the $z$ integration must be performed from $-\infty$ to $+\infty$
for the shift theorem to be applicable \cite{shift}.

Inserting complete set of $|p_z>$ states (using $\int dp_z |p_z><p_z|=1$)
as appropriate we find
\bea
&& S^{(1)}=\frac{i}{2(2\pi)^2} \sum_{j=1}^4 \int_0^\infty \frac{ds}{s} \int d^2x_T \int d^2p_T
e^{is(p_T^2+m^2+i\epsilon)} \nonumber \\
&&[ \int_{-\infty}^{+\infty} dt <t|\int dp_z\int dq_z  
\int_{-\infty}^{+\infty} dz <z+\frac{i}{E(t)}\frac{d}{dt}|p_z> \nonumber \\
&& <p_z| e^{is[-e^2E^2(t) z^2+\hat{p}_z^2-i\lambda_j E(t)]}|q_z> <q_z|z+\frac{i}{E(t)}\frac{d}{dt}>|t> - \int dt \int dz ~\frac{1}{4\pi s}].
\eea
Using $<z|p_z> = \frac{1}{\sqrt{2\pi}} e^{izp_z}$ we find
\bea
&& S^{(1)}
=\frac{i}{2(2\pi)^2} \sum_{j=1}^4 \int_0^\infty \frac{ds}{s} \int d^2x_T\int d^2p_T
e^{is(p_T^2+m^2+i\epsilon)} [ F_j(s) - \int dt \int dz~ \frac{1}{4 \pi s}]
\label{12}
\eea
where
\bea
&& F_j(s)=\frac{1}{(2\pi)} \int_{-\infty}^{+\infty} dt <t|\int dp_z\int dq_z
\int_{-\infty}^{+\infty} dz~ e^{izp_z}e^{-\frac{1}{E(t)}\frac{d}{dt}p_z}<p_z| 
 e^{is[-e^2E^2(t) z^2+\hat{p}_z^2-i\lambda_j E(t)]}|q_z>  \nonumber \\
 && e^{\frac{1}{E(t)}\frac{d}{dt} q_z} e^{-izq_z}|t>.
\label{gn}
\eea
It can be seen that the exponential $e^{-\frac{1}{E(t)}\frac{d}{dt}p_z}$ 
contains the derivative $\frac{d}{dt}$ which operates on
$<p_z|e^{is[-e^2E^2(t) z^2+\hat{p}_z^2-i\lambda_j E(t)]}|q_z> e^{\frac{1}{E(t)}\frac{d}{dt} q_z}$ 
hence we can not move $e^{-\frac{1}{E(t)}\frac{d}{dt}p_z}$ to right.

We insert complete sets of $|z>$ states and $|p_0> $states using the completeness relations:
\be
1= \int_{-\infty}^ {\infty} dz |z> < z|; ~~~    1= \int_{-\infty}^ {\infty} dp_0 |p_0> < p_0|
\ee
to find
\bea
&& F_j(s)=\frac{1}{(2\pi)} \int_{-\infty}^{+\infty} dt \int dp_0 \int dp'_0 \int dp''_0 \int dp'''_0 \int dp_z\int dq_z \int dz_1 \int dz_2
\int_{-\infty}^{+\infty} dz \nonumber \\
&& <t|p_0><p_0|e^{izp_z}e^{-\frac{1}{E(t)}\frac{d}{dt}p_z}|p'_0><p'_0|<p_z|z_1> 
<z_1| e^{is[-e^2E^2(t) z^2+\hat{p}_z^2-i\lambda_j E(t)]}|z_2><z_2|q_z>|p''_0> \nonumber \\
&& <p''_0|e^{\frac{1}{E(t)}\frac{d}{dt} q_z} e^{-izq_z}|p'''_0><p'''_0|t> \nonumber \\
&& =\frac{1}{(2\pi)^3} \int_{-\infty}^{+\infty} dt \int dp_0 \int dp'_0 \int dp''_0 \int dp'''_0 \int dp_z\int dq_z \int dz_1 \int dz_2
\int_{-\infty}^{+\infty} dz \nonumber \\
&& e^{itp_0} e^{izp_z}<p_0|e^{-\frac{1}{E(t)}\frac{d}{dt}p_z}|p'_0>e^{-iz_1p_z} <p'_0|
<z_1| e^{is[-e^2E^2(t) z^2+\hat{p}_z^2-i\lambda_j E(t)]}|z_2> |p''_0> \nonumber \\
&& e^{iz_2q_z} <p''_0|e^{\frac{1}{E(t)}\frac{d}{dt} q_z} |p'''_0> e^{-izq_z}e^{-itp'''_0}.
\label{gn5}
\eea

The matrix element, $<z_1| e^{is[-e^2E^2(t) z^2+\hat{p}_z^2-i\lambda_j E(t)]}|z_2>$ 
can be evaluated in terms of the eigenstates of the Harmonic Oscillator with
time dependent frequency. The Hamiltonian 
\be
H= \frac{1}{2} [p_z^2 + \omega^2(t) z^2]
\ee
can be diagonalized at every time $t$ in terms of 
time dependent creation and annihilation operators 
\bea
 b(t)= 
\sqrt{\frac{\omega(t)}{2}} ~z +~ \frac{i}{\sqrt{2 \omega(t)}}~ p_z~, 
~~~~~~~~~~b^\dag (t)=
\sqrt{\frac{\omega(t)}{2}} ~z -~ \frac{i}{\sqrt{2 \omega(t)}}~ p_z. 
\label{cran}
\eea
Hence we find
\be
H = \omega(t) [ b^\dag(t) b(t) + \frac{1}{2}]. 
\ee
The eigenvalues of the Hamiltonian are just
\be
\lambda_n = \omega(t) [n+\frac{1}{2}], ~~~~n= 0, 1,2, \cdots.
\ee
and the normalized eigenstates  $<z|n_t>$  are the usual 
harmonic oscillator ones with $\omega \rightarrow \omega(t)$.
\bea
[{\hat p}_z^2 + \omega^2(t)z^2]|n_t>=(2n+1)\omega(t) |n_t>
\eea
\bea
<z|n_t>=\psi_n(z)=(\frac{\omega(t)}{\pi})^{1/4}\frac{1}{(2^nn!)^{1/2}}
H_n(z \sqrt{\omega(t)})e^{-\frac{\omega(t)}{2}z^2}.
\label{hmt}
\eea
Since the creation and annihilation operators ($b^\dag (t)$ and $b(t)$)
in eq. (\ref{cran}) are constructed from $z$ and $p_z$ operators
the eigen state $|n_t>$ of the number operator 
${\hat n}(t)$(=$b^\dag (t) b(t)$) commutes with any function $f(t)$
which is independent of $z$ and $p_z$. Therefore we find
\bea
[{\hat p}_z^2 + \omega^2(t)z^2 +\lambda_j E(t) ]|n_t>=[(2n+1)\omega(t) +\lambda_j E(t)] |n_t>.
\eea
By using the orthonormality relation of Hermite polynomials we find
\bea
\int dz  |\psi_n(z)|^2=\int dz |<n_t|z>|^2 = \int dz (
\frac{\omega(t)}{\pi})^{1/2}\frac{1}{(2^nn!)}|H_n(z 
\sqrt{\omega(t)})|^2e^{-\omega(t)z^2}=1
\label{nor}
\eea
and
\bea
&& \int dz \psi^{*}_n(z) \psi_m(z) = \int dz <n_t|z><z|m_t> \nonumber \\
&& = \int dz (\frac{\omega(t)}{\pi})^{1/2}\frac{1}{(2^nn!)} 
H^{*}_n(z \sqrt{\omega(t)})H_m(z \sqrt{\omega(t)})e^{-\omega(t)z^2}
=\delta_{nm}.
\label{orth}
\eea
By using $\int dz |z><z| =1$ in the above equation we find
\bea
<n_t|m_t>=\delta_{nm}.
\label{o1}
\eea
The completeness condition is
\bea
\sum_n |n_t><n_t|=1. 
\label{c1}
\eea
In our case the harmonic oscillator $[-e^2E^2(t) z^2+\hat{p}_z^2]$ which appears
in eq. (\ref{gn5}) has imaginary \cite{itzy,gouranga} frequency $\omega(t) = ieE(t)$.
Inserting complete set of harmonic oscillator states $|n_t>$
($\sum_n |n_t><n_t|=1$) in eq. (\ref{gn5}) we find
\bea
&& F_j(s)=
\frac{1}{(2\pi)^3} 
\sum_n \sum_m \int_{-\infty}^{+\infty} dt \int dp_0 \int dp'_0 \int dp''_0 \int dp'''_0 \int dp_z\int dq_z \int dz_1 \int dz_2
\int_{-\infty}^{+\infty} dz \nonumber \\
&& e^{itp_0} e^{izp_z}<p_0|e^{-\frac{1}{E(t)}\frac{d}{dt}p_z}|p'_0>e^{-iz_1p_z} <p'_0|
<z_1|n_t> <n_t|e^{is[-e^2E^2(t) z^2+\hat{p}_z^2 - i\lambda_j E(t)]}|m_t> \nonumber \\
&& <m_t|z_2> |p''_0> e^{iz_2q_z} <p''_0|e^{\frac{1}{E(t)}\frac{d}{dt} q_z} |p'''_0> e^{-izq_z}e^{-itp'''_0} \nonumber \\
&& =
\frac{1}{(2\pi)^3} 
\sum_n  \int_{-\infty}^{+\infty} dt \int dp_0 \int dp'_0 \int dp''_0 \int dp'''_0 \int dp_z\int dq_z \int dz_1 \int dz_2
\int_{-\infty}^{+\infty} dz \nonumber \\
&& e^{itp_0} e^{izp_z}<p_0|e^{-\frac{1}{E(t)}\frac{d}{dt}p_z}|p'_0>e^{-iz_1p_z} <p'_0|
<z_1|n_t> e^{-s[eE(t)(2n+1)-\lambda_j E(t)]}<n_t|z_2> |p''_0> \nonumber \\
&& e^{iz_2q_z} <p''_0|e^{\frac{1}{E(t)}\frac{d}{dt} q_z} |p'''_0> e^{-izq_z}e^{-itp'''_0} \nonumber \\
&& =
\frac{1}{(2\pi)^3} 
\sum_n  \int_{-\infty}^{+\infty} dt \int dp_0 \int dp'_0 \int dp''_0 \int dp'''_0 \int dp_z\int dq_z \int dz_1 \int dz_2
\int_{-\infty}^{+\infty} dz \nonumber \\
&& e^{itp_0} e^{iz(p_z-q_z)}<p_0|e^{-\frac{1}{E(t)}\frac{d}{dt}p_z}|p'_0>e^{-iz_1p_z} <p'_0|
<z_1|n_t>e^{s\lambda_j E(t)} e^{-seE(t) (2n+1)}<n_t|z_2> |p''_0> \nonumber \\
&& e^{iz_2q_z} <p''_0|e^{\frac{1}{E(t)}\frac{d}{dt} q_z} |p'''_0> e^{-itp'''_0}.
\label{gn7}
\eea
As advocated earlier 
the $z$ integration must be performed from $-\infty$ to $+\infty$
for the shift theorem to be applicable \cite{shift}.
Performing  the $z$ integration we get
 \bea
\int_{-\infty}^{+\infty} dz e^{iz(p_z-q_z)} = 2 \pi \delta(p_z-q_z)
\eea
and we find
\bea
&&F_j(s) =
\frac{1}{(2\pi)^2} 
\sum_n  \int_{-\infty}^{+\infty} dt \int dp_0 \int dp'_0 \int dp''_0 \int dp'''_0 \int dp_z \int dz_1 \int dz_2 e^{itp_0} 
<p_0|e^{-\frac{1}{E(t)}\frac{d}{dt}p_z}|p'_0> \nonumber \\
&& e^{-iz_1p_z} <p'_0| <z_1|n_t>  e^{s\lambda_j E(t)} e^{-seE(t) (2n+1)} <n_t|z_2> |p''_0> e^{iz_2p_z} <p''_0|e^{\frac{1}{E(t)}\frac{d}{dt} p_z} |p'''_0> e^{-itp'''_0}. \nonumber \\
\label{gn8}
\eea
Inserting further complete sets of $|t>$ states we obtain
\bea
&&F_j(s)=
\frac{1}{(2\pi)^2} 
\sum_n  \int_{-\infty}^{+\infty} dt \int dt_1 \int dt_2 
\int dp_0 \int dp'_0 \int dp''_0 \int dp'''_0 \int dp_z \int dz_1 \int dz_2 
 e^{itp_0} \nonumber \\
 && <p_0|e^{-\frac{1}{E(t)}\frac{d}{dt}p_z}|p'_0> e^{-iz_1p_z} <p'_0|t_1>
<t_1|<z_1|n_t> e^{s\lambda_j E(t)} e^{-seE(t) (2n+1)}<n_t|z_2> \nonumber \\
&& |t_2><t_2|p''_0>  e^{iz_2p_z} <p''_0|e^{\frac{1}{E(t)}\frac{d}{dt} p_z} |p'''_0> e^{-itp'''_0} \nonumber \\
&& =
\frac{1}{(2\pi)^3} 
\sum_n  \int_{-\infty}^{+\infty} dt \int dt_1 \int dt_2 
\int dp_0 \int dp'_0 \int dp''_0 \int dp'''_0 \int dp_z \int dz_1 \int dz_2 \nonumber \\
&& e^{itp_0} <p_0|e^{-\frac{1}{E(t)}\frac{d}{dt}p_z}|p'_0>e^{-iz_1p_z} e^{-it_1p'_0}
<t_1|<z_1|n_t> e^{s\lambda_j E(t)} e^{-seE(t) (2n+1)}<n_t|z_2> \nonumber \\
&& |t_2>e^{it_2p''_0} e^{iz_2p_z} <p''_0|e^{\frac{1}{E(t)}\frac{d}{dt} p_z} |p'''_0> e^{-itp'''_0}.
\label{gn9}
\eea
 Since $<n_t|z>$ does not contain any time derivative $\frac{d}{dt}$ operator 
 (see eq. (\ref{hmt})) we find (by using $<t_1|t_2>=\delta(t_1-t_2)$)
 \bea
&&F_j(s) =
\frac{1}{(2\pi)^3} 
\sum_n  \int_{-\infty}^{+\infty} dt \int dt_1 
\int dp_0 \int dp'_0 \int dp''_0 \int dp'''_0 \int dp_z \int dz_1 \int dz_2 \nonumber \\
&& e^{itp_0} <p_0|e^{-\frac{1}{E(t)}\frac{d}{dt}p_z}|p'_0>e^{-iz_1p_z} e^{-it_1p'_0}
<z_1|n_{t_1}>e^{s\lambda_j E(t_1)}  e^{-seE(t_1) (2n+1)}<n_{t_1}|z_2> e^{it_1p''_0} \nonumber \\
&& e^{iz_2p_z} <p''_0|e^{\frac{1}{E(t)}\frac{d}{dt} p_z} |p'''_0> e^{-itp'''_0}.
\label{gn10}
\eea
Since $<p_0|e^{-\frac{1}{E(t)}\frac{d}{dt}p_z}|p'_0>$ and
$<p''_0|e^{\frac{1}{E(t)}\frac{d}{dt} p_z} |p'''_0>$ are independent of the time derivative
operator $\frac{d}{dt}$ (they only depend on c-numbers $p_0$, $p'_0$, $p''_0$ and $p'''_0$)
we can take $<p''_0|e^{\frac{1}{E(t)}\frac{d}{dt} p_z} |p'''_0>$ and $e^{-itp'''_0}$ to the left. We find
 \bea
&&F_j(s) =
\frac{1}{(2\pi)^3} 
\sum_n  \int_{-\infty}^{+\infty} dt \int dt_1 
\int dp_0 \int dp'_0 \int dp''_0 \int dp'''_0 \int dp_z \int dz_1 \int dz_2 \nonumber \\
&& e^{it(p_0-p'''_0)} <p''_0|e^{\frac{1}{E(t)}\frac{d}{dt} p_z} |p'''_0><p_0|e^{-\frac{1}{E(t)}\frac{d}{dt}p_z}|p'_0>e^{-iz_1p_z} e^{-it_1p'_0}
<z_1|n_{t_1}>e^{s\lambda_j E(t_1)} e^{-seE(t_1) (2n+1)} \nonumber \\
&& <n_{t_1}|z_2> e^{it_1p''_0} e^{iz_2p_z}.
\label{gn12}
\eea
It can also be shown that $<p_0|e^{-\frac{1}{E(t)}\frac{d}{dt}p_z}|p'_0>$ and
$<p''_0|e^{\frac{1}{E(t)}\frac{d}{dt} p_z} |p'''_0>$ are independent of
$t$. This can be shown as follows:
\bea
&& <p_0|f(t)\frac{d}{dt}|p'_0>= \int dt' \int dt'' \int dp''''_0 <p_0|t'> <t'|f(t)|t''><t''|p''''_0><p''''_0|\frac{d}{dt}|p'_0> \nonumber \\
&&= \int dt' \int dt'' \int dp''''_0 e^{-it' p_0}~ \delta(t'-t'') f(t'')
e^{it'' p''''_0}~
ip'_0~ \delta(p''''_0-p'_0)=ip'_0 \int dt'  ~e^{-it'(p_0-p'_0)}f(t') \nonumber \\
\eea
which is independent of $t$. Hence we can easily integrate over $t$ in eq.
(\ref{gn12}).

Integrating over $t$ in eq. (\ref{gn12}) by using $\int_{-\infty}^{+\infty} dt e^{it(p_0-p'''_0)} = 2 \pi \delta(p_0-p'''_0)$, we find
 \bea
&&F_j(s) =
\frac{1}{(2\pi)^2} 
\sum_n  \int dt_1 
\int dp_0 \int dp'_0 \int dp''_0  \int dp_z \int dz_1 \int dz_2 <p''_0|e^{\frac{1}{E(t)}\frac{d}{dt} p_z} |p_0> <p_0|e^{-\frac{1}{E(t)}\frac{d}{dt}p_z}|p'_0> \nonumber \\
&& e^{-iz_1p_z} e^{-it_1p'_0}
<z_1|n_{t_1}> e^{s\lambda_j E(t_1)} e^{-seE(t_1) (2n+1)} <n_{t_1}|z_2> e^{it_1p''_0} e^{iz_2p_z}.
\label{gn13}
\eea
Since $\int dp_0 |p_0><p_0| =1$ and 
$\int dp_z e^{ip_z(z_1-z_2)}=2\pi \delta(z_1-z_2)$ we find
\bea
&&F_j(s)=
\frac{1}{(2\pi)^2} 
\sum_n  \int dt_1 
 \int dp'_0 \int dp_z \int dz_1 \int dz_2  e^{-iz_1p_z}
<z_1|n_{t_1}> e^{s\lambda_j E(t_1)} e^{-seE(t_1) (2n+1)} \nonumber \\
&& <n_{t_1}|z_2>  e^{iz_2p_z}  = \frac{1}{(2\pi)} \sum_n  \int dt  \int dp'_0  \int dz 
|<z|n_{t}>|^2 ~e^{s\lambda_j E(t)} e^{-seE(t) (2n+1)}.
\label{gn15}
\eea
Since the harmonic oscillator wave function is normalized (see eq. (\ref{nor})):
\bea
\int dz |<z|n_t>|^2=1
\eea
we find
\bea
F_j(s)= \frac{1}{(2\pi)} 
\sum_n  \int dt  \int dp_0 e^{s\lambda_j E(t)} e^{-seE(t) (2n+1)} = \frac{1}{(2\pi)} 
\int dt  \int dp_0 \frac{e^{s\lambda_j E(t)}}{{2 \rm sinh}(seE(t))}.
\label{gn16}
\eea
Using the Lorentz force equation
\bea
dp_\mu = eF_{\mu \nu}dx^\nu
\eea
we find (when $E(t)$ is along $z$-axis, eq. (\ref{7}))
\bea
dp_0=eE(t) dz.
\label{dt}
\eea
Hence 
\bea
F_j(s)=
\frac{1}{(4\pi)} 
\int dt  \int dz  \frac{eE(t) e^{s\lambda_j E(t)}}{{ \rm sinh}(seE(t))}.
\label{gn17}
\eea
Using the above expression for $F$ in eq. (\ref{12}) the effective action becomes
\bea
S^{(1)}=\frac{i}{32\pi^3} \sum_{j=1}^4  \int_0^\infty \frac{ds}{s} \int d^4x \int d^2p_T
e^{is(p_T^2+m^2+i\epsilon)} [ \frac{eE(t) e^{s\lambda_j E(t)}}{{ \rm sinh}(seE(t))}- ~ \frac{1}{s}]
\label{122}
\eea
Using the values of $\lambda_j$ (the eigen values of the Dirac matrix $\gamma^0 \gamma^3$ 
from eq. (\ref{dig})) we find
\bea
S^{(1)}=
\frac{i}{8 \pi^3} 
\int_0^\infty \frac{ds}{s} \int d^4x \int d^2p_T
e^{is(p_T^2+m^2+i\epsilon)} 
[eE(t) ~{\rm coth}(seE(t))~-~\frac{1}{s}]. 
\label{12ff}
\eea
The imaginary part of the above effective action gives real particle pair
production. Using the series expansion 
\bea
\frac{1}{{\rm sinh} x}=\frac{1}{x} + 2x \sum_{n=1}^\infty \frac{(-1)^n}{\pi^2 n^2 +x^2}
\eea
the s-contour integration is straight forward \cite{gouranga,itzy,schw}. 
Performing the s-contour integration around the pole $s=\frac{in\pi}{|eE(t)|}$
we find for the probability  $W$ of pair production (which is twice
the  imaginary part of the effective action)
\bea
W=2 {\rm Im} S^{(1)}=
\frac{1}{4\pi^3} 
\sum_{n=1}^\infty \frac{1}{n} \int d^4x \int d^2p_T |eE(t)|
e^{-n\pi \frac{p_T^2+m^2}{|eE(t)|}}.
\label{14ff}
\eea
Therefore the probability of producing a fermion (or antifermion) per unit volume 
per unit time with transverse momentum $p_T$ from arbitrary time dependent electric field 
$E(t)$ is given by 
\bea
\frac{dW}{d^4xd^2p_T}=
-\frac{|eE(t)|}{4\pi^3} {\rm Log}[1-e^{-\pi \frac{p_T^2+m^2}{|eE(t)|}}]
\label{dwf}
\eea
which reproduces eq. (\ref{dw}).

To conclude we have obtained an exact one-loop non-perturbative 
result for the probability of fermion-antifermion pair production per unit time per unit 
volume per unit transverse momentum (of the fermion or antifermion)
from the arbitrary time dependent electric field $E(t)$ via Schwinger mechanism. We have 
found that the exact non-perturbative result is independent of all the 
time derivatives $\frac{d^nE(t)}{dt^n}$, where $n=1,2,....\infty$.

The quantum back reaction problem for the Electric Field is studied in 
\cite{cooper}.  There it is shown that the exact numerical solution is consistent with the solution of the semi-classical transport equation
having a source term for pair production which utilizes Schwinger's
constant electric field $E$  result but with the replacement 
$E~\rightarrow E(t)$.  
This result  seems to be consistent
with our finding here. However in those simulations the Schwinger source term was modified by
Bose Enhancement effects for boson pair production and Pauli-blocking 
effects for fermion pair production \cite{cooper,tr1}. These Pauli-blocking 
effects are not present in our calculation which needs to be better understood.
Particle production from arbitrary time (and space) dependent background 
fields is important in early universe dynamics  and in the  production and 
equilibration of the quark-gluon plasma at RHIC and LHC \cite{cooper,tr1,grr1}. 

\acknowledgments
We thank George Sterman and Peter van Nieuwenhuizen for 
discussions. This work was supported in part by the National Science 
Foundation, grants PHY-0354776 and PHY-0345822. Fred Cooper would like 
to thank Harvard University for its hospitality during the writing of 
this paper.

\end{document}